# Laser induced rotation of trapped chiral and achiral nematic droplets

Marjan Mosallaeipour[†,*], Yashodhan Hatwalne[*], N.V. Madhusudana[*], and Sharath Ananthamurthy[†]

*†Department of Physics, Jnanabharathi Campus, Bangalore University, Bangalore 560056, INDIA*

[*] *Raman Research Institute, C.V. Raman Ave., Sadashivanagar, Bangalore 560012, INDIA*

marjan@rri.res.in, yhat@rri.res.in, nvmadhu@rri.res.in, asharath@gmail.com

We study the response of optically trapped achiral and chiralised nematic liquid crystal droplets to linear as well as circular polarised light. We find that there is internal dissipation in rotating achiral nematic droplets trapped in glycerine. We also demonstrate that some chiralised droplets rotate under linearly polarised light. The best fit to our data on chiralised droplets indicates that rotational frequency of these droplets with radius $R$ is approximately proportional to $1/R^2$, rather than to $1/R^3$.

# 1. Introduction

Transfer of photon momentum to nematic droplets has been studied since Juodkazis *et al* [1] demonstrated rotation of nematic droplets optically trapped in heavy water. They also constructed a fast optical switch using this rotation. Because of the possible application of liquid crystal droplets as micro-rotors and as optical switches, this field of study has gained increasing importance in recent years.

Several mechanisms have been suggested for the transfer of angular momentum of circular polarized light to nematic slabs and droplets. The relative contributions of these mechanisms (such as wave-plate behaviour, absorption, nonlinear effects and scattering processes) were compared by Wood *et al* [2], who showed that wave-plate behaviour [3] of the droplet is the dominant mechanism in generating rotation. In the wave-plate mechanism, the change in the angular momentum of light after passage through the birefringent nematic droplet results in a corresponding torque on the droplet.

The rotational dynamics of nematic droplets (assumed to be spheres) have been analysed for the following two simple cases: (*i*) the droplet rotates as a *rigid body* without any internal dynamics and (*ii*) there is flow in surrounding fluid as well as director rotation, the rotational dynamics of both assumed to be *different but uniform*. The theoretical details of droplet rotational dynamics in both these cases have been discussed by Manzo *et al* [4, 5].

If the droplet is assumed to be a rigid body of radius R (case (*i*) above), the angular frequency of rotation under circularly polarized light can be estimated to be [3]

$$\Omega = P[1 - \cos k 2R\Delta n]/(\omega D), \qquad (1)$$

where $\Delta n = n_e - n_o$ is the birefringence of the medium, $D$ is the drag coefficient ($D = 8\pi\eta R^3$ for a sphere of radius $R$ in a medium of viscosity $\eta$), $P$ is the trapping power of the laser, $\omega$ is the angular frequency of the laser light, and $k$ is the wavenumber. We note that $\Omega$ is proportional to $1/R^3$, which is characteristic of rigid body rotation; the viscous dissipation occurs only in the external fluid in which the droplet is immersed. Manzo *et al* [4, 5] experimentally verified that the measured rotational frequency of a nematic droplet of radius $R$ in water (which has low shear viscosity as compared to the nematic) follows Stokes' law; the angular frequency of the droplets is proportional to $1/R^3$. We discuss case (*ii*) in relation to our experiments on nematic droplets trapped in glycerine in Section 4.

In this work we describe our experimental studies on the rotation of droplets of 4-pentyl-4′-cyanobiphenyl (5CB of Merck) which exhibits the nematic phase at the ambient temperature. We first carry out studies of the rotation of nematic droplets suspended in (*i*) water, and (*ii*) glycerine. We then study the rotation of chiralised nematic droplets under circular *as well as linear polarized light*.

Our results are as follows: (a) We verify the rigid body-like behaviour of optically trapped nematic droplets (trapped in water) under circularly polarized light, (b) we also verify that the approximation (*ii*) (see Eqn. (2)) describes the dynamics of nematic droplets in glycerine, which is a highly viscous medium, (c) we observe rotation of optically trapped chiral nematic droplets suspended in water under circularly polarized light and show that these droplets do not follow dynamics described by either (*i*) or (*ii*) above. Further, we show that the droplet angular frequency of the droplets is approximately proportional to $1/R^2$, rather than to $1/R^3$, (d) we also show that some trapped chiral droplets can rotate under linearly polarized light [6], whereas some do not, in agreement with the results reported in [7]. Droplets which rotate under linearly polarized light do not change the sense of rotation under circularly polarized light if the sense of circular polarization is switched (from left- to right-). However, droplets which do not rotate under linearly polarized light do change the sense of rotation if the sense of circular polarization is switched. The details of our results for both nematic and chiral droplets are given in the sections below.

## 2. Experimental details

For performing the optical rotation experiments an optical trap setup similar to the one described in [8] was used. A 100x oil-immersion (Olympus IX70) microscope objective (NA=1.4) was used to focus the input laser beam. The liquid crystal droplet was imaged between a polarizer and a crossed analyzer using white light. The trapping beam was a linearly polarized Gaussian laser beam at 1064nm (Coherent, USA), with output power of 100mW. Using a power meter (THORLABS, USA) we measured the laser power after transmission through the objective to be around 1.5mW. For producing a circularly polarized beam a quarter wave plate (Newport) was used. For imaging, white light was passed through the sample from above onto a CCD (with video camera module, model CE N50).

Chiral nematic liquid crystals were prepared by doping 5CB with cholesteryl nonanoate at different concentrations. The mixture was heated to isotropic temperature of nonanoate (around 80° C) and stirred. The pitch of the mixture with a concentration of 17wt% of nonanoate was found to be about 3.4μm by measurements on a finger print texture [9]. The cholesteric helix of this mixture is left-handed [10]. A suspension of cholesteric liquid crystal droplets was created by dispersion of a small amount of the mixture in double distilled water followed by sonication in an ultrasonic bath; many cholesteric liquid crystal droplets in the 1-30 μm diameter range were formed.

Following a similar procedure pure 5CB nematic liquid crystal droplets were formed in water and glycerine. The size of droplets was controlled by changing the time of sonication. A small amount of the suspension was directly placed on a cover slip, the other side of which made contact with the immersion oil of the objective. The top side of the sample was left open. The experiments were carried out at room temperature. Tracing the texture of the droplet allows for direct measurement of the droplet rotation (see movies 1 and 2 in the electronic supplementary material).

## 3. Rotation of nematic droplets suspended in water

Nematic droplets of 5CB in water have a bipolar structure with the director aligned parallel to the nematic- water interface [11]. The viscosity coefficient of the nematic material is much larger than that of water (for 5CB at room temperature the rotational viscosity is about 70cP [12], whereas for water the shear viscosity is about 1cP). The rotation is slow enough that the droplets retain their spherical shape while rotating under an external torque. Thus it is natural to expect that the nematic droplets rotate as rigid bodies under circularly polarized light as has been already noted previously [4,5]. The rotation of nematic liquid crystal droplets under circularly polarized light is uniform and the sense of rotation of the droplet changes with the change of sense of polarization. Our results are summarized in Figure 1.

**4. Rotation of nematic droplets in a high viscosity medium**

Nematic droplets made of 5CB were trapped in glycerine, a high viscosity medium (shear viscosity around 600 cP at room temperature [13]), and with a trap strength which is lower as compared with that in water as the refractive index of glycerine $n_g$=1.47 is greater than water. As in water, the structure of the droplets in glycerine is bipolar, with tangential anchoring of the director at the nematic-glycerine interface. Again, rotation of droplets was observed under circularly polarized light. The viscosity of nematic droplets is lower than that of the surrounding medium and viscous dissipation occurs predominantly within the droplet (see the discussion following equation 2). Figure 2 shows the rotational frequency of nematic droplets in glycerine as a function of droplet radius. We note that the rotation rate of droplets in glycerine is much higher than that expected for a rigid body. It is thus appropriate to compare the experimental results with the analysis corresponding to case (*ii*) in the Introduction.

In case (*ii*), the uniform but different angular velocities of the surrounding fluid as well as that of the director have to be taken into account. The maximum angular frequency of the director rotation $\Omega_n$ and that of the surrounding fluid $\Omega_v$ are given by [5]:

$$\Omega_n = \frac{3P}{4\pi\omega R^3}\left(\frac{1}{6\eta} + \frac{1}{\gamma_1}\right), \quad (2A)$$

$$\Omega_v = \frac{P}{8\pi\eta\,\omega R^3}, \quad (2B)$$

where $\gamma_1$ is the rotational viscosity of the nematic [12] and $\eta$ is the shear viscosity of the surrounding medium. We note that both the angular frequencies are proportional to $1/R^3$ and the rotation can still be considered "Stokesian", albeit with different "effective viscosities", or with different "effective radii". This is a direct consequence of the assumptions leading to the derivation of equations (2A) and (2B). Note that the effective viscosity has a structure akin to a parallel combination (in analogy with resistors in parallel) of the rotational viscosity of the nematic and that of the surrounding medium. Thus the lower viscosity (that of the nematic) dominates. Thus it is reasonable to expect that the dissipation occurs predominantly within the nematic droplet. This view is further corroborated by Rayleigh's principle of minimum dissipation in the theory of dissipative systems [14].

Equation 2A fits the experimental data as shown in Figure 2. As expected, the direction of rotation of the droplet changes with the sense of polarization.

## 5. Measurements on chiralized droplets

Figure 3 demonstrates the rotation of a chiral droplet under linearly polarized light. In order to clearly visualize the rotation of a large droplet, a mixture with concentration 1.5wt% of cholesterol nonanoate and 33 μm pitch length was used.

The director field of a chiral droplet depends upon the radius R of the droplet, the pitch $p$ and the surface anchoring [15]. A mixture with 17wt% of cholesterol nonanoate has a pitch of 3.4 μm, and droplets in the range of about 2-3 μm radii rotate under linear polarized light. Droplets with larger radii do not rotate under linearly polarized light but do so under circularly polarized light. The textures of drops that rotate as well as those that do not under linearly polarized light are shown in Figure 4.

Figure 5 shows our data for chiral droplets. We find that droplets that rotate under linearly polarized light also rotate under circularly polarized light. However on changing the sense of polarization, the sense of rotation of these droplets remains *unchanged*. For other droplets, the sense of rotation changes with the sense of circularly polarized light. The diameter to pitch ratio for chiral droplets which rotate under linearly polarized light was found to lie roughly between 1 to 2 for mixtures with 17 wt% of cholesteryl nonanoate, and between 0.15 to 0.45 in mixtures with 1.5 wt% of cholesteryl nonanoate.

In drops which do rotate under the action of linearly polarized light, there are slight differences in the frequencies of rotation of a given droplet when the laser beam has linear, left circular and right circular polarizations (Figure 6).

Gleeson and co-workers [16] have reported the rotation of optically trapped chiralised droplets under plane polarized light. These authors speculate that the chiral torque generated by optical activity of the medium followed by elastic relaxation of the photoinduced change in the structure is a possible mechanism for the rotation [7]. We were unaware of the experiments of Gleeson and co-workers at the time when our initial observations on chiralised droplets were made [6].

In Figure 5 the unconstrained nonlinear least-square fit of the data to the form $f = aR^n$ gives $a = 0.4$ with $n = -1.9$. We note that the rotation of chiralised droplets is jerky under linear as well as circular polarised light. Further, there are two kinds of droplets as discussed above. In view of this we performed an independent set of experiments and found that the best fit to the data yields $a = 0.33$ with $n = -2.2$ (see the supplementary material). This leads us to believe that the rotational frequency of chiralised droplets is approximately proportional to $1/R^2$ (rather than $1/R^3$, as is the case for nematic droplets suspended in water or glycerine).

A fit of the form $f = aR^{-3}$ to the data for chiral droplets gives $a = 0.94$ whereas a = 5 for nematic droplets in water. A comparison of the data in Figures 1 and 5 shows that the rate of rotation is much lower for chiral droplets than that of nematic droplets. The novel result pertaining to the low rate of rotation of chiral droplets needs a detailed analysis. The analysis of Freise *et al* [3] can be extended to a helical stack of birefringent plates that mimic the cholesteric structure. However, this will not lead to rotation of the structure in a linearly polarized laser beam. *Therefore, it is necessary to take into account the light-*

*induced deformation of the helical structure for the complete analysis of the problem*. We are currently developing such a model.

**6. Summary and conclusions**

In conclusion, we have studied the rotation of laser trapped achiral and chiralized nematic droplets under circularly polarized light. Further, we have shown that in a highly viscous medium like glycerine there is internal dissipation within the droplets in contrast to those suspended in water. We have demonstrated the rotation of some chiralized droplets under linearly polarized light, in agreement with the observations of [16]. We also find that the rotational frequency of chiral droplets is approximately proportional to $1/R^2$, in contrast to that of nematic droplets for which it is proportional to $1/R^3$. We suggest a possible mechanism for the novel result pertaining to the rotation of chiralised droplets.

Acknowledgements
M.Mosallaeipour acknowledges financial aid and the use of laboratory facilities from the Raman Research Institute.

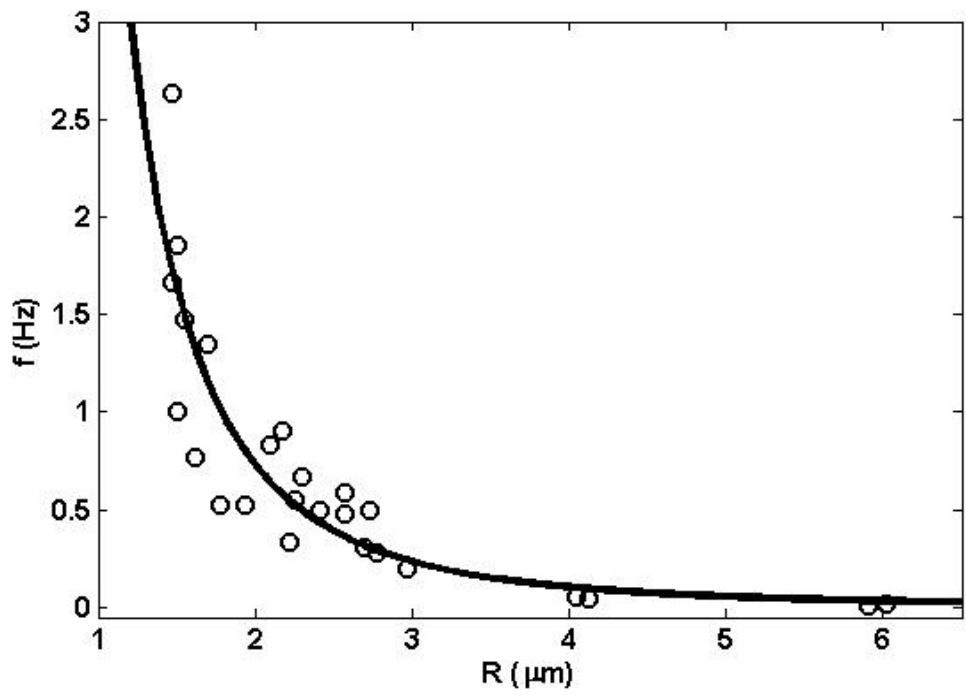

Figure 1. Rotational frequency f of nematic 5CB droplets in water under circularly polarized light. Solid line is the theoretical prediction for rigid body rotation of the droplet.

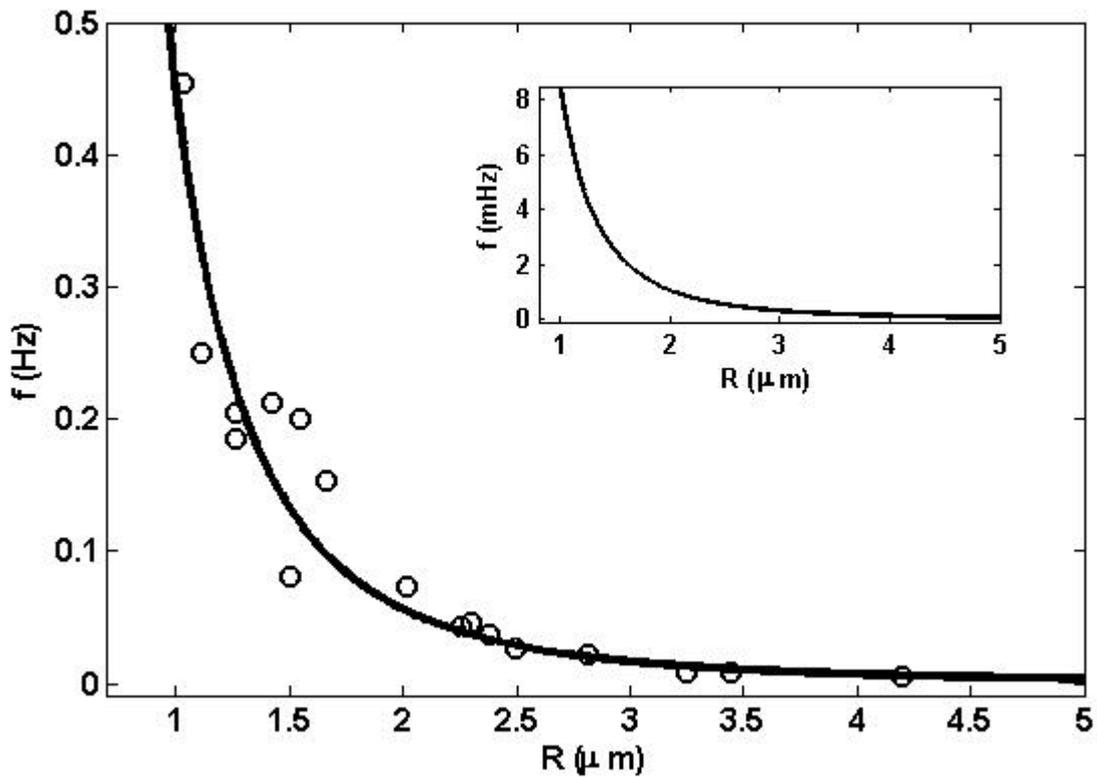

Figure 2. Rotational frequency f of nematic droplets in glycerine under circularly polarized light. The solid line is the theoretical prediction according to equation 2A. The inset shows the theoretical prediction for a rigid body rotation of the droplets which the rotational frequency is in the mHz range.

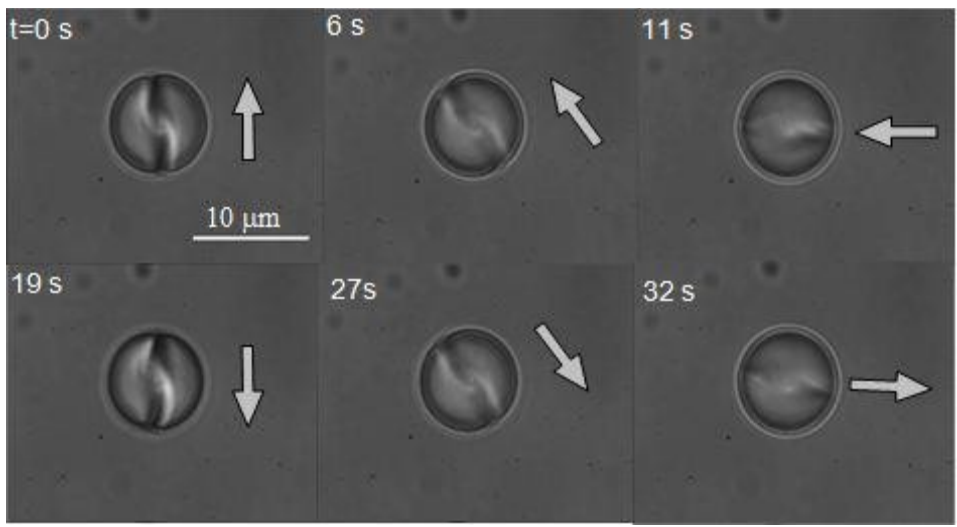

Figure 3. Time evolution of rotation for a trapped chiral droplet under linear polarized light. The droplet diameter is 8.5 µm and has 1.5wt% concentration of cholesteryl nonanoate, the chiral pitch $p \approx 33$ µm. Images are taken without analyser for visible light.

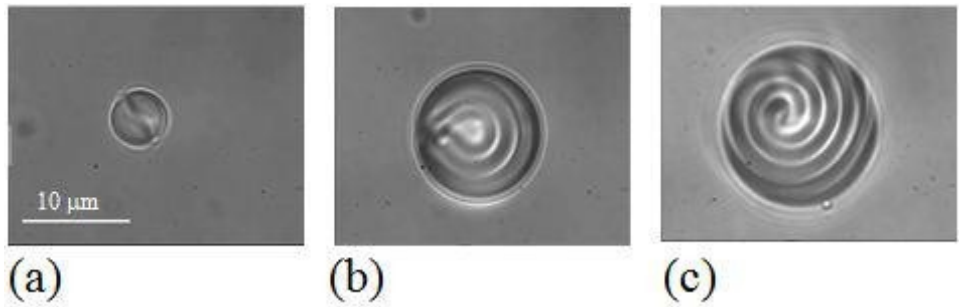

Figure 4. Rotational dynamics of chiralised drops with 17wt % of cholesteryl nonanoate (a) Drop with radius R=2.5μm rotates under linear as well as circularly polarized light, whereas drops with (b) R=6μm, and (c) R=7.5μm, rotate under circularly polarized light but not under linearly polarized light. The trapping power is 1.5mw. In (a), (b) and (c) the chiral pitch $p \approx 3.4$ μm.

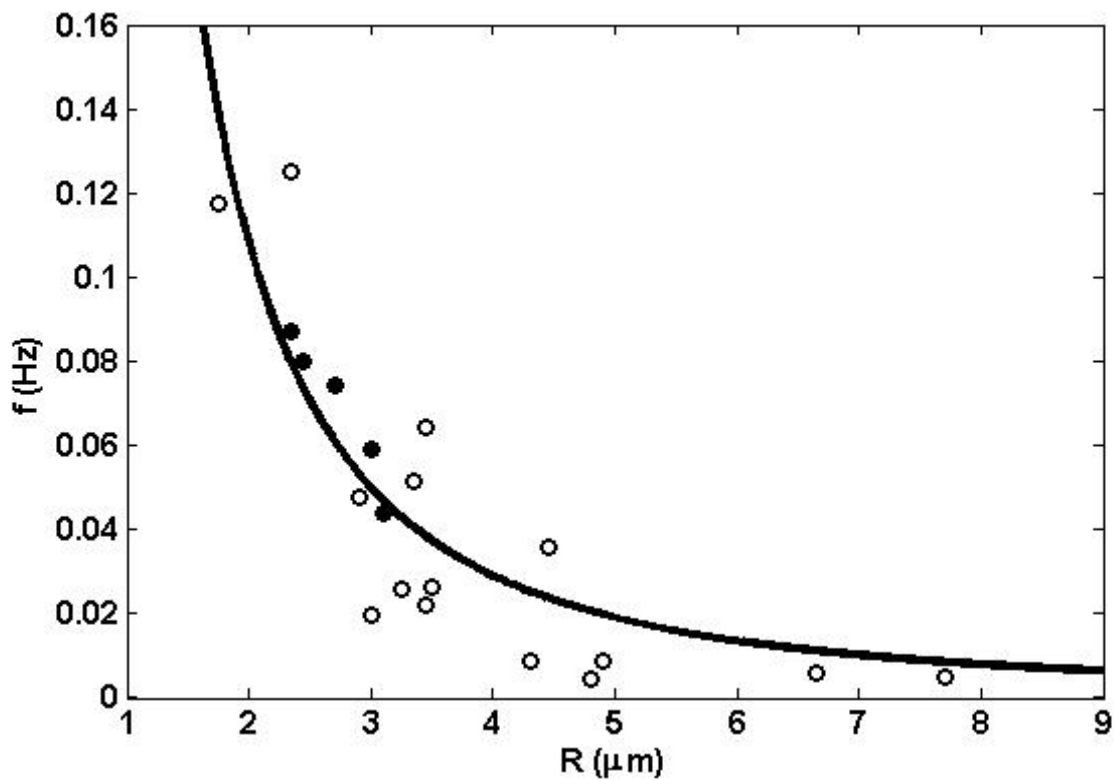

Figure 5. Filled circles correspond to droplets that rotate under linearly as well as circularly polarized light. The other droplets rotate only under circularly polarized light, and all the data correspond to left circularly polarized light. Solid line is the best fit to the data, and shows that f is approximately proportional to $1/R^2$ (see the main text).

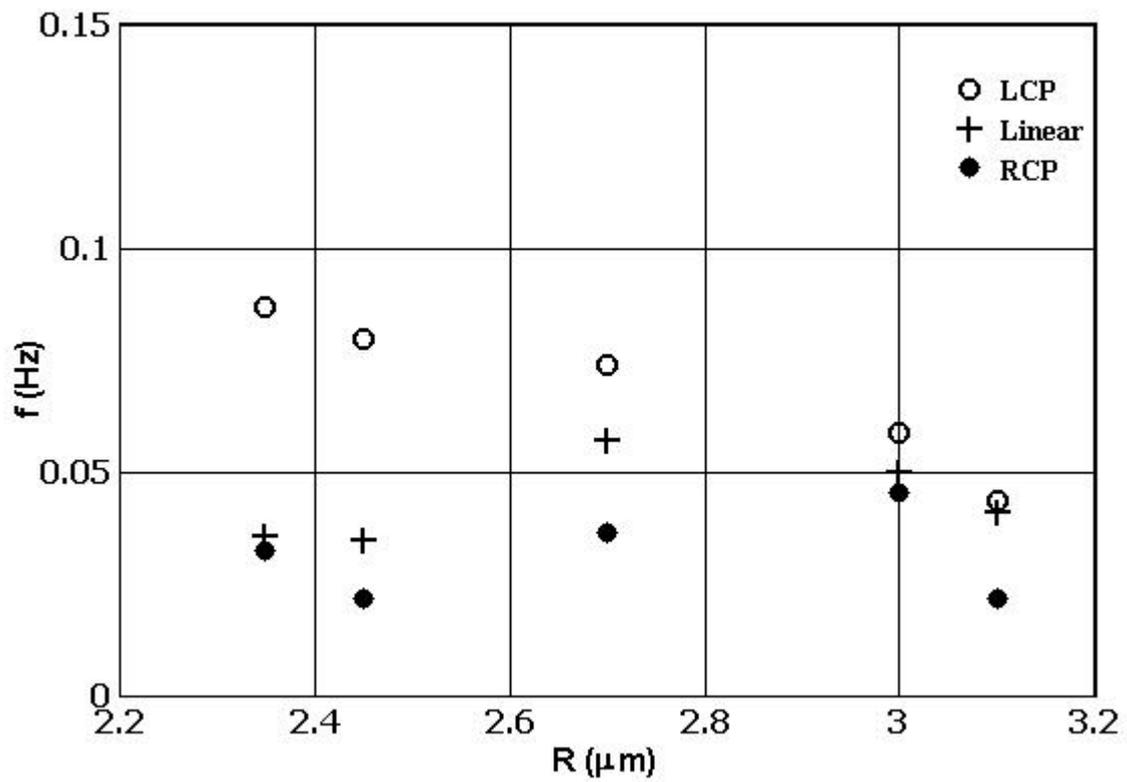

Figure 6. Rotational frequency f of chiral droplets with 3.4 μm pitch under linearly and circularly polarized light (RCP: right- circular polarization, LCP: left- circular polarization).